\documentclass[aip,jmp,reprint,onecolumn]{revtex4-1}
\usepackage[latin1]{inputenc}
\usepackage{amsmath}
\usepackage{amsfonts}
\usepackage{amssymb}
\usepackage{amsthm}
\usepackage{color}
\usepackage[english]{babel}
\usepackage{indentfirst}
\usepackage{amssymb}
\usepackage{mathrsfs}
\usepackage{latexsym}
\usepackage{graphicx}
\usepackage{slashed}
\newcommand{\beq}{\begin{eqnarray}}
\newcommand{\eeq}{\end{eqnarray}}
\newcommand{\bc}{\begin{center}}
\newcommand{\ec}{\end{center}}

\newtheoremstyle{query}%
{}{}
{\color{red}}
{}
{\sffamily\bfseries}{:}{12pt}
{}

\begin{document}

\title{Aharonov--Bohm scattering for relativistic particles in (3 + 1)-dimensional noncommutative space with spin dependence}

\author{C. A. Stechhahn}

\affiliation{Faculdades Integradas Campos Salles, Rua Nossa Senhora da Lapa, 270/284 -- Lapa -- CEP 05072-000 -- São Paulo, SP, Brazil}

\begin{abstract}
We study the effects of  noncommutativity of  spacetime with mixed spatial and spin degrees of freedom
in a relativistic context. Using the Dirac equation in $(3+1)$ dimensions and in a symmetric gauge, we calculate the invariant amplitude
for a small magnetic field flux. The parameter $\theta$ that characterizes the noncommutativity here is not constant, and the model
preserves Lorentz symmetry.
A comparison is made with  scattering in the context of canonical noncommutativity.
\end{abstract}

\maketitle
 
\section{Introduction} 

 In recent years, several aspects of noncommutative quantum theories have been investigated (see Refs.~\onlinecite{RevModPhys.73.977} for a review). In these studies,
noncommutativity is  considered to be a fundamental description  of spacetime at the Planck scale
 \cite{CommMathPhys.172.187} or as a particular limit of a quantum theory of gravity.\cite{JHEP.09.032} 
 
 Noncommutative coordinates were first proposed as long ago as 1947 in an attempt to remove the ultraviolet
divergences that plague perturbative quantum field theories,\cite{PhysRev.71.38} although related ideas were discussed
even earlier than that. \cite{NuclPhys.B108.30} However, the idea was abandoned
after the great success of the renormalization program. 

With the advent of string theory, new interest in  noncommutative (NC) space emerged. \cite{PhysRev.D84.045002,PhysRev.D86.105035,JPhys.A.46.245303,ModPhysLett.A.29.1450048,EurPhysJ.C74.3101,PhysLett.A378.3509,PhysLett.B743.184,PhysLett.B761.462,JHEP.01.049,PhysLett.B766.181,PhysLett.A382.3317,EurPhysJ.C79.300,UkrJPhys.64.131}
It has been shown that the dynamics of open strings in the presence of an antisymmetric $B$-field can be described in certain limits
by a gauge theory on a noncommutative spacetime, \cite{JHEP.09.032} namely, one deformed by the Moyal product (whose definition will be given
shortly). 

One well-known physical example of noncommutativity
of coordinates appears in quantum mechanics. The Landau problem \cite{PhysLett.B718.1475} in which a charged particle moves in a
strong constant magnetic field that is perpendicular to the plane of motion leads to a  noncommutative space when it is projected
to a lower Landau  level. 

At distances of the order of the
Planck length,  owing to the generation of intense gravitational fields, the measurement of coordinates loses all meaning, as does the concept of ``point.'' This suggests the introduction of position operators that do not commute:
\begin{equation} \label{eq1.1}
[\hat{x}_\mu,\hat{x}_\nu]=i\theta_{\mu\nu},  
\end{equation}
 where $\theta_{\mu\nu}$ is a constant antisymmetric matrix of dimension (length)$^2$ and the hatted quantities
are  position operators in NC space. The nonlocality inherent to Eq.~(\ref{eq1.1}) leads to the appearance of 
several significant effects, such as  ultraviolet/infrared (UV/IR) mixing. \cite{JHEP.03.035} In this context, another
interesting aspect of NC spacetime is the violation of unitarity and causality when $\theta_{0i}\ne 0$. \cite{NuclPhys.B591.265,PhysLett.B533.178,EurPhysJ.C25.469,JHEP10.2004.072,JHEP.06.044,JHEP.05.040,PhysRev.D69.065008} However, in all these situations,  $\theta_{\mu\nu}$ is constant, i.e,
we have  canonical noncommutativity. It is therefore appropriate to consider novel cases where, for example, the
noncommutativity is spin-dependent \cite{PhysLett.B680.384}. We should emphasize here that the NC  Aharonov--Bohm  (AB) effect 
has been studied  in both the quantum mechanical \cite{PhysRev.D76.085008} and   field theory  \cite{PhysRev.D70.085005,PhysRev.D71.107701} contexts. More general situations may also be considered where $\theta_{\mu\nu}$ 
is position-dependent, i.e., a new operator. In this sense,  Eq.~(\ref{eq1.1}) may be taken as a first approximation
to a more general setting where the commutator of coordinates itself is not a constant operator.
\cite{AnnPhys.260.224,PhysLett.B.555.83,EurPhysJ.54.325,PhysRev.D79.125011} In Refs.~\onlinecite{PhysLett.B718.1475} and \onlinecite{JPhys.A.MathTheor.43.285301}, the physical
meaning of spin noncommutativity is discussed in terms of  a nonstandard Heisenberg algebra. 

In the present work, we shall consider a relativistic quantum mechanical model where
the spin dependence is incorporated into the Pauli--Lubanski vector. We calculate the scattering amplitude for relativistic particles
in a symmetric gauge. The aim of this work is to study  this novel kind of noncommutativity and its physical
implications in contrast to the canonical case.

\section{The Aharonov--Bohm effect: symmetric gauge and  spin noncommutativity}

We consider the AB effect, with an infinitely thin solenoid lying along the $z$ axis in an NC space
and with coordinates $\hat{x}^\mu$ satisfying the algebra
\begin{equation} \label{CR}
\left[\hat{x}^\mu,\hat{x}^\nu\right]=-i\theta\epsilon^{\mu\nu\rho\sigma} S_{\rho\sigma}+i\frac{\theta^2}{2}\epsilon^{\mu\nu\rho\sigma}W_\rho p_\sigma,
\end{equation}
 where 
\begin{equation} \label{PL}
W^\mu =\frac{1}{2}\epsilon^{\mu\nu\rho\sigma} S_{\rho\sigma}=\frac{1}{2}\gamma^5 \sigma^{\mu\nu}\partial_\nu ,
\end{equation}
 is the Pauli--Lubanski pseudovector, $S_{\rho\sigma}=\sigma_{\rho\sigma}/2$ is the spin operator, and $\sigma^{\mu\nu}=\frac{i}{2}[\gamma^\mu,\gamma^\nu]$. 
Whenever necessary, we assume the Pauli representation for the gamma matrices:
\begin{equation} \label{S1}
 \gamma^0=\begin{pmatrix}
 \mathbf{I} & 0 \\
0 & -\mathbf{I} 
 \end{pmatrix}, \qquad \gamma^i=\begin{pmatrix}
 0 & \sigma^i  \\
-\sigma^i & 0 
 \end{pmatrix},
\end{equation}
 where the  spacetime metric is $\eta^{\mu\nu}= \mathrm{diag}(1,-1,-1,-1)$ and  $\sigma^i$ are the Pauli matrices. The boost and
rotation generators are \cite{AnIntroductiontoQuantumFieldTheory.41}
\begin{equation} \label{S2}
 S^{0i}=\frac{i}{4}[\gamma^0,\gamma^i],\qquad S^{ij}=\frac{i}{4}[\gamma^i,\gamma^i],
\end{equation} 
 so  the antisymmetric $S^{ij}=\sigma^{ij}/2$ and $\gamma^5=i\gamma^0\gamma^1\gamma^2\gamma^3$.
The algebra in Eq.~(\ref{CR}) can be implemented through the following definitions: 
\begin{equation}\begin{aligned} \label{PauliLub}
 x^\mu&\rightarrow\hat{x}^\mu = x^\mu \mathbf{1}+\theta W^\mu,
\\[3pt]
p^\mu&\rightarrow\hat{p}^\mu = p^\mu,
\end{aligned}\end{equation}
 where $x^\mu$ and $p^\mu$ satisfy the usual Heisenberg algebra.

We write the NC Dirac equation in a form that ensures a Hermitian Lagrangian density:\cite{PhysLett.B718.1475}
\begin{equation} \label{NCDiracEquation}
 \left\{-i\gamma^\mu \partial_\mu + M +\frac{e}{2}\left[A_\mu (x)\gamma^\mu\star + A_\mu(x)\star\gamma^\mu\right]\right\}\psi(x)=0,
\end{equation}
where the ``star operation'' (Moyal product) represents the following action of a Weyl-ordered operator $f(\hat{x}^\mu)$ on a spinor $\psi(x)$:
\begin{equation} \label{star} 
 f(\hat{x}^\mu)\star \psi=f(x^\mu\mathbf{1}+\theta\gamma^5 \sigma^{\mu\nu}\partial_\nu)\psi
 =f(x^\mu) \exp\!\left(\theta\overleftarrow{\partial}_{\mu} \gamma^5 \sigma^{\mu\nu}\overrightarrow{\partial}_{_\nu}\right)\psi.
\end{equation}
 We adopt the symmetric ordering shown in Eq.~(\ref{NCDiracEquation})  because the operator
$A_\mu(\hat{x}^\mu)$ has a matrix structure that does not commute
with $\gamma^\mu$. In the relativistic case, the NC Dirac equation describes a spin-1/2
particle in an external electromagnetic field in the presence of  spin noncommutativity.

In ordinary relativistic quantum mechanics, the Dirac equation for a particle of mass $M$ and  charge $-e$ in an external magnetic
field reads
\begin{equation} \label{DiracEquation}
 \left[\alpha_i\left(p_i-eA_i\right)+\beta M\right]\psi(j,x)=i\partial_t\psi(j,x),
\end{equation}
 where  $\alpha_i=\gamma^0\gamma^i$ and $\beta=\gamma^0$.
 The magnetic field is derived from a vector
potential $(A^0=A^3=0)$ such that
\begin{equation} \label{eA}
 eA_i=-\frac{e\Phi}{2\pi}\frac{\epsilon_{ij}x_j}{\rho^2}=-\alpha \frac{\epsilon_{ij}x_j}{\rho^2},
\end{equation}
where $\Phi$ is the magnetic flux, $\epsilon_{ij}$ is the Levi-Civita symbol 
(normalized as $\epsilon_{12}=1$), 
 and $\rho=\sqrt{x_1^2+x_2^2}$\,. This is the gauge potential (symmetric gauge) in an ordinary commutative space, and it corresponds
 to $\vec{B}=\Phi\delta(\rho)\hat{k}$. 
 
Thus, after performing the star product
in the NC Dirac equation in the AB potential, up to first order in $\theta$, Eq.~(\ref{NCDiracEquation}) reads
\begin{equation} \label{DiracEq14}
 \left[-i\gamma^0\gamma^i\partial_i +M\gamma^0 +e\gamma^0\gamma^i A_i(x) 
+\frac{e\theta}{2}\left(\partial_k A_i\right)\gamma^0\left\{\gamma^i, \gamma^5 \sigma^{kj}\right\}\partial_j \right]\psi(x)=i\partial_t\psi(x).
\end{equation}
We are using natural units, such that $\hbar =c=1$, and $e<0$ for the electron charge.
A straightforward calculation in the symmetric gauge (\ref{eA}) leads to the total Hamiltonian $H$ as the sum of
\begin{align}  
H_0&=-i\gamma^0\gamma^i\partial_i +M\gamma^0 -\frac{e\Phi}{2\pi\rho^2}\epsilon_{ij}x_j\gamma^0\gamma^i \label{Ho}\\
 \intertext{and}
H_\mathrm{int}&=\frac{e\theta}{2} [\left(\partial_1 A_i\right)\gamma^0\left\{\gamma^i, \gamma^5\sigma^{12}\right\}\partial_2
+
 \left(\partial_2 A_i\right)\gamma^0\left\{\gamma^i, \gamma^5\sigma^{21}\right\}\partial_1 ]\psi
 =i\partial_t\psi.\label{Hint}
\end{align}
 After all the $\gamma$ products have been carried out and the
 $x_i$ and $\partial_i$ terms have been properly grouped, $H_\mathrm{int}$ reduces to
\begin{align} \label{Hint2}
 H_\mathrm{int}={}&\alpha\theta \left\{\frac{2x_1 x_2}{\rho^4}\begin{pmatrix}
 -i\sigma^2 & 0 \\
0 & -i\sigma^2 
 \end{pmatrix}\partial_2
 +\left(\frac{1}{\rho^2}-\frac{2x_1^2}{\rho^4}\right)\begin{pmatrix}
 i\sigma^1 & 0 \\
0 & i\sigma^1
 \end{pmatrix}\partial_2\right.\nonumber\\[6pt]
&\hspace{16pt}\left.+\left(\frac{1}{\rho^2}-\frac{2x_2^2}{\rho^4}\right)\begin{pmatrix}
 -i\sigma^2 & 0 \\
0 & -i\sigma^2
 \end{pmatrix}\partial_1
+\frac{2x_1 x_2}{\rho^4}\begin{pmatrix}
 i\sigma^1 & 0 \\
0 & i\sigma^1
 \end{pmatrix}\partial_1\right\}
\end{align}
 (for details, see Appendices~\ref{appB} and \ref{appC}). After some algebra, this interaction Hamiltonian may be written as 
\begin{equation} \label{Hint3}
 H_\mathrm{int}(\rho,\varphi)=-\frac{\alpha\theta}{\rho^3} 
\begin{pmatrix}
 0 & e^{-i\varphi}z^{+}_{ij} & 0 & 0 \\
 e^{i\varphi}z^{-}_{ij} & 0 & 0 & 0 \\
0 & 0 & 0 & e^{-i\varphi}z^{+}_{ij} \\
0 & 0 & e^{i\varphi}z^{-}_{ij} & 0 
 \end{pmatrix},
 \end{equation}
  where  
 \begin{equation} \label{z}
  z^{\pm}_{ij}=\left(i\epsilon_{ij}\pm \delta_{ij}\right)x_i\partial_j.
 \end{equation}
   
  The Dirac spinor solution of the Dirac equation in an AB potential, Eq.~(\ref{DiracEquation}),
may be written as \cite{PhysRev.D53.2178}
\begin{equation} \label{psi}
 \psi(j,x)=\frac{1}{2\pi}\frac{1}{\sqrt{2E_p}}e^{-iE_pt+ip_3z} e^{im\varphi}e^{i\pi|\ell|/2}\begin{pmatrix}
 \chi \\[3pt]
  \zeta 
  \end{pmatrix},
\end{equation}
where $m$ is the integer part of the parameter $\alpha=e\Phi/2\pi$  in Eq.~(\ref{eA}): $\alpha=m+\delta$, with $0<\delta<1$.
The two-component eigenvectors are given by 
\begin{align}
 \chi&=\frac{1}{\sqrt{2s}}
\begin{pmatrix}
\sqrt{E_p+sM)}\sqrt{s+1}J_{|\ell-\delta|}(p^{\perp}\rho)e^{i\ell\varphi} \\[4pt]
  i\epsilon_3\epsilon_{|\ell-\delta|}\sqrt{E_p-sM)}\sqrt{s-1}J_{|\ell-\delta|+\epsilon(\ell-\delta)}(p^{\perp}\rho)e^{i(\ell+1)\varphi}
  \end{pmatrix} \label{u}\\
\intertext{and} 
\zeta&=\frac{1}{\sqrt{2s}}\begin{pmatrix}
 \epsilon_3 \sqrt{E_p+sM)}\sqrt{s-1}J_{|\ell-\delta|}(p^{\perp}\rho)e^{i\ell\varphi} \\[4pt]
  i\epsilon_{|\ell-\delta|}\sqrt{E_p-sM}\sqrt{s+1}J_{|\ell-\delta|+\epsilon(\ell-\delta)}(p^{\perp}\rho)e^{i(\ell+1)\varphi}
  \end{pmatrix},\label{v}
\end{align}
where $J_{|\ell-\delta|}$ is the Bessel function of the first kind, $\epsilon_3:=\mathrm{sign}(sp_3)$, $\ell$ is an integer,
and $s$ is the eigenvalue of the equation $\hat{S}\psi=s\psi$. In what follows, we take $\mathrm{sign}(sp_3)=1$. Since
$p^2=E_p^2-M^2$,
we have 
\begin{equation} \label{pp}
 (p^{\perp})^2=p^2-p_3^2,\qquad s=\pm\sqrt{1+\frac{p_3^2}{M^2}}\,.
\end{equation}

\begin{figure}[h!]
\includegraphics[width=0.5\columnwidth]{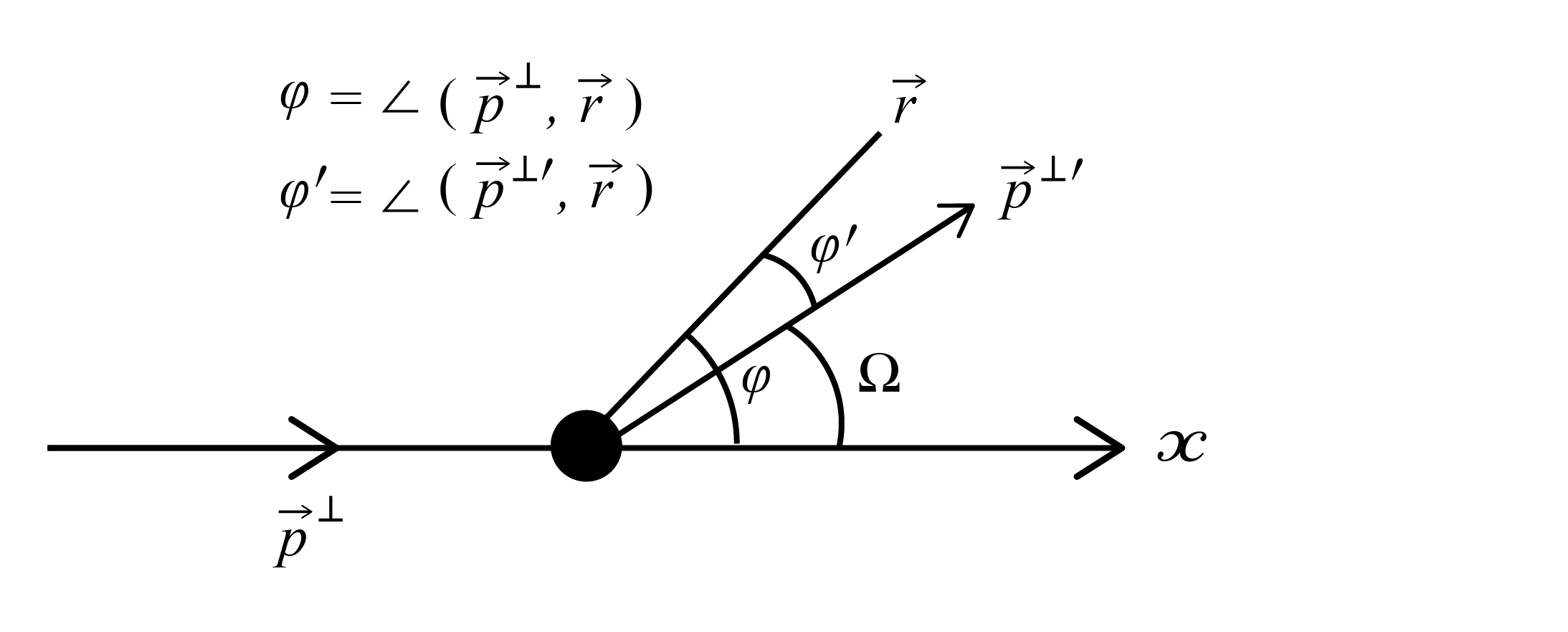}
\caption{Geometry of the Aharonov--Bohm effect. A magnetic field $\vec{B}$ is confined in a straight solenoid along the
$z$ axis, which contains a magnetic flux $\Phi$. The cross section of the solenoid is shown here as the black circle and there is no magnetic
field outside it. Relativistic charged particles with energy $E_p$ and mass $M$ are incident from the negative $x$ direction. They
are scattered by the solenoid, but cannot penetrate  it. \label{fig1}}
\end{figure}

The first-order Born amplitude for the scattering of a Dirac particle induced by spin noncommutativity is then
\begin{align} \label{SA}
 S_{fi}^{(1)}&=-i\int dt \,\langle \psi_f|H_\mathrm{int}|\psi_i\rangle
 \nonumber
\\
&=-i\int dt \,d^3x\, \psi_f^\dagger(x) H_\mathrm{int}\psi_i(x) e^{i(E_f-E_i)t} e^{i(p^\prime _3-p_3)z}
\nonumber
\\
&=-i(2\pi)^2 i\delta(E_f-E_i)\delta(p^\prime _3-p_3)f_{fi}^{(1)}.
 \end{align}
This result emerges from the $t$ and $z$ integrations corresponding to  energy and momentum conservation
 in the $z$ direction. Thus, we are left with calculation of
 the following integral:
 \begin{equation}
  f_{fi}^{(1)}=\int \rho\, d\rho\, d\varphi \,\psi_f^\dagger(p^{\perp}\rho) H_\mathrm{int} \psi_i(p^{\perp}\rho).
 \end{equation}
 This integration is over all space, and $\psi$ is the Dirac spinor given by Eq.~(\ref{psi}). Furthermore,
we introduce the following notation for some terms in the spinor components:
\begin{align} \label{Ep+}
 \sqrt{E_p+ sM}&:= \surd{E_{p+}},\qquad\sqrt{s+ 1}:=\surd{s_{+}},\\[4pt]
 \label{Ep-}
 \sqrt{E_p- sM}&:= \surd{E_{p-}},\qquad\sqrt{s- 1}:=\surd{s_{-}},
\end{align}
\begin{align} \label{epsilon}
\epsilon_\ell&:=\epsilon_{|\ell-\delta|}=\begin{cases}
1, & \ell>\delta,\\
-1 ,& \ell<\delta,
\end{cases}\\[6pt] 
J_{\nu_{i}}&:=\begin{cases}
J_{|\ell-\delta|}, & i=1,\\
J_{|\ell-\delta|-\epsilon(\ell-\delta)},& i=2,
\end{cases}
\end{align}
 and similarly for their Hermitian adjoints $\epsilon_{\tilde{\ell}}$ and $J_{\tilde{\nu}_{i}}$ ($\tilde{\ell}$ is
the azimuthal quantum number of the outgoing wave). With these considerations,
the invariant amplitude can be written as
\begin{equation} \label{Hint4}
S_{fi}^{(1)}=-\frac{ie\theta \Phi}{8\pi E_p s}\int_0^\infty \int_0^{2\pi} \frac{1}{\rho^2}\, d\rho\, d\varphi
 \mathbf{v}^\dagger \begin{pmatrix}
0 & e^{-i\ell\varphi}z^+_{ij} & 0 & 0 \\[2pt]
e^{i\ell\varphi}z^-_{ij} & 0 & 0 & 0 \\[2pt]
0 & 0 & 0 & e^{-i\ell\varphi}z^+_{ij} \\[2pt]
0 & 0 & e^{i\ell\varphi}z^-_{ij} & 0
\end{pmatrix} \mathbf{u},
\end{equation}
where  
\begin{align}
 \mathbf{u}&=\begin{pmatrix}
\surd{E_{p+}}\surd{s_{+}}J_{\nu_1}e^{i\ell\varphi} \\[3pt]
i\epsilon_\ell \surd{E_{p-}} \surd{s_{-}}J_{\nu_2}e^{i(\ell+1)\varphi} \\[3pt]
 \surd{E_{p+}}\surd{s_{-}}J_{\nu_1}e^{i\ell\varphi} \\[3pt]
i\epsilon_\ell \surd{E_{p-}}\surd{s_{+}}J_{\nu_2}e^{i(\ell+1)\varphi}
\end{pmatrix}\\
\intertext{[which, up to constants, is the spinor in Eq.~(\ref{psi})] and}
\mathbf{v}&=\begin{pmatrix}
\surd{E_{p+}}\surd{s_{+}}J_{\tilde{\nu}_1}e^{i\tilde{\ell}\tilde{\varphi}} \\[3pt]
i\epsilon_{\tilde{\ell}} \surd{E_{p-}} \surd{s_{-}}J_{\tilde{\nu}_2}e^{i(\tilde{\ell}+1)\tilde{\varphi}} \\[3pt]
 \surd{E_{p+}}\surd{s_{-}}J_{\tilde{\nu}_1}e^{i\tilde{\ell}\tilde{\varphi}} \\[3pt]
i\epsilon_{\tilde{\ell}} \surd{E_{p-}}\surd{s_{+}}J_{\tilde{\nu}_2}e^{i(\tilde{\ell}+1)\tilde{\varphi}}
\end{pmatrix},
\end{align}
and $\dagger$ indicates the Hermitian conjugate. Here, $\tilde{\varphi}=\varphi-\Omega$, where $\Omega$ is the scattering angle (see Fig.~\ref{fig1}).

Equation~(\ref{Hint4}) involves spatial derivatives given by the $z$ operators defined in Eq.~(\ref{z}). After multiplying all
the matrices, we get
\begin{align}
S_{fi}^{(1)}=\frac{e\theta \Phi}{4\pi E_p s}\mathcal{W}\int_0^\infty \int_0^{2\pi}  d\rho\, d\varphi
 &\left\{\epsilon_\ell \frac{1}{\rho^2} [(\ell+1)J_{\tilde{\nu}_1}J_{\nu_2}+J_{\tilde{\nu}_1}J_{\nu_2}^\prime p^\perp \rho ]
 e^{-i\tilde{\ell}\tilde{\varphi}+i\ell\varphi}\right.
 \nonumber
 \\[4pt]
 &\hspace{6pt}\left.-\epsilon_{\tilde{\ell}} \frac{1}{\rho^2} [\ell J_{\tilde{\nu}_2}J_{\nu_1}-J_{\tilde{\nu}_2}J_{\nu_1}^\prime p^\perp \rho ]
 e^{-i(\tilde{\ell}+1)\tilde{\varphi}+i(\ell+1)\varphi}\right\},
\end{align}
where $\mathcal{W}=\surd{E_{p+}}\surd{E_{p-}}\surd{s_{+}}\surd{s_{-}}$.
The above expression can be cast as
\begin{align}
 S_{fi}^{(1)}=\frac{e\theta \Phi}{4\pi E_p s}\mathcal{W}\int_0^\infty \int_0^{2\pi}  d\rho\, d\varphi
  & \left\{  \left[\frac{\epsilon_\ell}{\rho^2}(\ell+1)J_{\tilde{\nu}_1}J_{\nu_2}+\frac{\epsilon_\ell}{\rho}J_{\tilde{\nu}_1}J_{\nu_2}^\prime p^\perp \right]
 e^{i(\ell-\tilde{\ell})\varphi+i\tilde{\ell}\Omega}\right.
 \nonumber
 \\[5pt]
 &\hspace{6pt}\left.+ \left(-\frac{\epsilon_{\tilde{\ell}}}{\rho^2}\ell J_{\tilde{\nu}_2}J_{\nu_1}+\frac{\epsilon_{\tilde{\ell}}}{\rho}J_{\tilde{\nu}_2}J_{\nu_1}^\prime
 p^\perp\right )e^{i(\ell-\tilde{\ell})\varphi+i(\tilde{\ell}+1)\Omega}  \right\},
\end{align}
 where the prime denotes differentiation with respect to the argument of the Bessel functions of the first kind, which are given explicitly 
in Eqs.~(\ref{u}) and (\ref{v}). We can write $\tilde{\varphi}$ in terms of the scattering angle $\Omega$, and in this way we can easily
perform the $\varphi$ integration, yielding
\begin{align} \label{Sfi1}
S_{fi}^{(1)}=\frac{e\theta \Phi}{2 E_p s}\mathcal{W}\lim_{\lambda\rightarrow 1} \int_0^\infty d\rho 
& \left\{ \left[\frac{\epsilon_\ell}{\rho^{2\lambda}}(\ell+1)J_{\nu_1}J_{\nu_2}+\frac{\epsilon_\ell}{2\rho^{2\lambda -1}} (J_{\nu_1}J_{{\nu_2}-1}-J_{\nu_1}J_{{\nu_2}+1}  )p^\perp  \right]e^{i\ell\Omega}\right.
 \nonumber
 \\[5pt]
 &\hspace{6pt}\left.+ \left[\frac{-\epsilon_\ell}{\rho^{2\lambda}}(\ell+1)J_{\nu_1}J_{\nu_2}+\frac{\epsilon_\ell}{2\rho^{2\lambda -1}} (J_{\nu_2}J_{{\nu_1}-1}-J_{\nu_2}J_{{\nu_1}+1}  )p^\perp \right ]e^{i(\ell+1)\Omega} \right \},
\end{align}
where we have introduced a parameter $\lambda$ so that, after all the integrations have been performed, we can investigate whether
its physical value $\lambda = 1$ can be taken. 
 The $\rho$ integrals can be performed \cite{GradshteynRyzhik} using the analytical continuation of the formula
\begin{equation} \label{Ffi2}
\int_0^\infty\frac{J_\nu(kr)J_\mu(kr)}{r^\sigma}dr=\frac{k^{\sigma-1}\Gamma(\sigma)\Gamma(A)}{2^{\sigma}\Gamma(B)\Gamma(C)\Gamma(D)},
\end{equation}
 where the  arguments in the gamma functions and the calculation method are shown explicitly  in Appendix~\ref{appA}. 

After performing all the integrations in Eq.~(\ref{Sfi1}), we obtain for the invariant amplitude
\begin{equation} \label{SAfinal}
 S_{fi}=
 \frac{e\theta \Phi p^{\perp}}{8 E_p}\frac{\mathcal{W}}{s} \sum_\ell  \frac{\epsilon_\ell[(2\ell-\delta+2)e^{i\ell\Omega}
 -(\ell+1)e^{i(\ell+1)\Omega}]}{(\ell-\delta)(\ell-\delta+1)}.
 \end{equation}
 After some manipulations, we find from  Eq.~(\ref{pp}) that 
\begin{equation}
 \frac{\mathcal{W}}{s}=\frac{p^{\perp}p_3}{\sqrt{M^2 +p_3^2}},
\end{equation}
 where $p^{\perp}=|\vec{p}^{\perp}|$ and the $z$ component of momentum are related to the relativistic energy by
$E_p^2 =(p^{\perp})^2 +p_3^2 +M^2$. 

\section{Conclusions}

In this paper, we have studied  spin noncommutativity for a relativistic system through  deformation of the spacetime 
coordinates $x^\mu$ into a nonstandard Heisenberg algebra given by Eq.~(\ref{CR}). 

In the nonrelativistic situation, the  AB scattering amplitude for spin-$1/2$ particles subject to this kind of noncommutativity  was
calculated in  Ref.~\onlinecite{PhysLett.B680.384}. Generalizing that study to the case of spin-$s$ relativistic  particles, we have calculated in $(3+1)$ dimensions the
scattering amplitude
in the AB symmetric gauge.

For canonical noncommutativity, the AB problem for a solenoid of finite radius was considered in
Ref.~\onlinecite{PhysRev.D66.045018}. In that
regime, the gauge potential was expanded in powers of the noncommutativity parameter and the magnetic flux $\Phi$. There, the first order in $\theta$
appeared multiplied by $\Phi^2$, and thus, since 
we are considering here a small magnetic flux, this term is  outside the scope of the present model. 

We have restricted
attention to  the potential as written in Eq.~(\ref{eA}). In our calculations, because of symmetry considerations, the energy and $p_3$ are
conserved.

From Eq.~(\ref{SAfinal}), we see that the scattering amplitude depends explicitly on $\delta$, the noninteger part of $\alpha$, the
term involving the flux. This is quite different  from the case of canonical noncommutativity
in $(2+1)$ dimensions, where $\delta$ is absorbed into $\alpha$. \cite{PhysRev.D76.085008}

The canonical NC algebra in Eq.~(\ref{eq1.1}) has generated many questions regarding violations of unitarity and causality.
This is primarily a consequence of the NC parameter $\theta$ being a constant matrix. However, the present model considering  spin noncommutativity
has the advantage over the canonical one in that it conserves Lorentz symmetry, since
the algebra in Eq.~(\ref{CR}) is written in a covariant way. In addition, there is no UV/IR
mixing; i.e., the limit $\theta\rightarrow 0$ is smooth.

The study of relativistic or even nonrelativistic quantum mechanics with NC coordinates
allows the examination of  phenomena that occur at very small scales and the exploration of
their physical consequences in a simple setting. However, it should be noted that  physical effects related
to quantum gravity will appear only in very high-energy processes. 

\begin{acknowledgments} 
I would like  to thank Professor M. Gomes for useful discussions and suggestions.
\end{acknowledgments}

\appendix
\section{Integrals of Bessel functions in Eq.~(\ref{Sfi1})\label{appA}}

The wave contributions for $\ell \le 0$ and $\ell >0$, given by the integrals of the double products of Bessel functions in Eq.~(\ref{Sfi1}), can be
calculated using the following result: 
\begin{equation} \label{Ffi2}
\int_0^\infty\frac{J_\nu(kr)J_\mu(kr)}{r^\sigma}\,dr=\frac{k^{\sigma-1}\Gamma(\sigma)\Gamma(A)}{2^{\sigma}\Gamma(B)\Gamma(C)\Gamma(D)},
\end{equation}
where
\begin{align} \label{2A}
2A&=\mu+\nu-\sigma+1,\\
 \label{2B}
2B&=-\nu+\mu+\sigma+1,\\
 \label{2C}
2C&=\mu+\nu+\sigma+1,\\
 \label{2D}
2D&=\nu-\mu+\sigma+1
\end{align}
is formally valid provided that the inequalities 
\begin{equation} \label{Ffi3}
\mathrm{Re}(\mu+\nu+1)>\mathrm{Re}(\sigma)>0
\end{equation}
are satisfied.

\subsection{The first integral in Eq.~(\ref{Sfi1})} 

The contribution for $\ell \le 0$ leads to the following integral:
\begin{equation} \label{B7}
\lim_{\lambda\rightarrow1}\int_0^{\infty}dr\,\frac{e^{i\ell\Omega}(\ell+1)}{r^{2\lambda}}J_{\alpha-\ell}(kr)J_{\alpha-\ell-1}(kr)
=\frac{e^{i\ell\Omega}(\ell+1)}{2}\frac{(\frac{1}{2}k)^{\sigma-1}\Gamma(A)}{\Gamma(B)\Gamma(C)\Gamma(D)},
\end{equation}
where $\sigma =2\lambda$ and
\begin{align} \label{2A}
2A&=\mu+\nu-\sigma+1=2\alpha-2\ell-2\lambda,\\
 \label{2B}
2B&=-\mu+\nu+\sigma+1=2\lambda,\\
 \label{2C}
2C&=\mu+\nu+\sigma+1=2\lambda+2\alpha-2\ell,\\
 \label{2D}
2D&=\mu-\nu+\sigma+1=2\lambda+2.
\end{align}

The condition (\ref{Ffi3})  (with $\lambda\rightarrow 1$) implies
\begin{equation} \label{Cond1}
\alpha-\ell+\alpha-\ell-1+1>2,
\end{equation}
i.e.,
\begin{equation} \label{Cond1}
\alpha>1+\ell,
\end{equation}

\noindent so that for $\ell < 0$ (and excluding the case $\ell =0$), we have $0<\alpha<1$. This kind of divergence can be eliminated by
not taking  $\lambda\rightarrow 1$ directly. Instead, we first perform  the integral and then take the limit. The integral in Eq.~(\ref{B7})
then becomes
\begin{equation} \label{B7a}
\lim_{\lambda\rightarrow1}\int_0^{\infty}dr\,\frac{e^{i\ell\Omega}(\ell+1)}{r^{2\lambda}}J_{\alpha-\ell}(kr)J_{\alpha-\ell-1}(kr)=\frac{e^{i\ell\Omega}(\ell+1)}{4}\frac{k\Gamma(\alpha-\ell-1)}{\Gamma(\alpha-\ell+1)\Gamma(2)},
\end{equation}
 where $\Gamma(\alpha-\ell+1)=(\alpha-\ell)(\alpha-\ell-1)\Gamma(\alpha-\ell-1)$.
The other integrals can be calculated following the same  approach. 

\section{The $z_{ij}$ operator in Eq.~(\ref{z})\label{appB}}

To obtain the expression for the $z_{ij}$ operator in Eq.~(\ref{z}), we have to use  expressions for the Pauli matrices $\sigma^i$ 
and remember that the $x_i$ ($i=1,2$) are in cylindrical coordinates. We have
\begin{align} 
H_\mathrm{int}=\frac{e\theta}{2}\frac{\Phi}{\pi} &\left\{\frac{2x_1 x_2}{\rho^4}\begin{pmatrix}
0 & -1 & 0 & 0 \\
1 & 0 & 0 & 0 \\ 
0 & 0 & 0 & -1 \\
0 & 0 & 1 & 0 
 \end{pmatrix}\partial_2 +\left(\frac{1}{\rho^2}-\frac{2x_1^2}{\rho^4}\right)\begin{pmatrix}
0 & i & 0 & 0 \\
i & 0 & 0 & 0 \\ 
0 & 0 & 0 & i \\
0 & 0 & i & 0 
 \end{pmatrix}\partial_2\right.
 \nonumber
 \\
 &\hspace{6pt}\left.+\left(\frac{1}{\rho^2}-\frac{2x_2^2}{\rho^4}\right)\begin{pmatrix}
0 & -1 & 0 & 0 \\
1 & 0 & 0 & 0 \\ 
0 & 0 & 0 & -1 \\
0 & 0 & 1 & 0 
 \end{pmatrix}\partial_1+\frac{2x_1 x_2}{\rho^4}\begin{pmatrix}
0 & i & 0 & 0 \\
i & 0 & 0 & 0 \\ 
0 & 0 & 0 & i \\
0 & 0 & i & 0 
 \end{pmatrix}\partial_1\right\},
\end{align}
where $\rho^2=x_1^2+x_2^2.$
The nonzero matrix elements are
\begin{align}
 a_{12}=a_{34}&= [-2x_1x_2 + i(x_2^2+x_1^2) ]\partial_2+ [(x_2^2-x_1^2)+2ix_1x_2  ]\partial_1
 \nonumber
 \\
 &= [-x_1x_2 + ix_2^2-x_1x_2-ix_1^2 ]\partial_2+ [ix_1x_2+x_2^2+ix_1 x_2-x_1^2  ]\partial_1
 \nonumber
 \\
 &= [-x_2(x_1-ix_2)-ix_1(x_1-ix_2) ]\partial_2+ [ix_2(x_1-ix_2)-x_1(x_1-ix_2)  ]\partial_1
 \nonumber
 \\
 &=\rho e^{-i\varphi} [(-x_2 - ix_1)\partial_2+(ix_2-x_1)\partial_1 ]
  \nonumber
 \\
 &=\rho e^{-i\varphi} [-i\epsilon_{ij}x_i\partial_j-\delta_{ij}x_i  \partial_j ]
 \nonumber
 \\
 &=-\rho e^{-i\varphi} [i\epsilon_{ij}+\delta_{ij} ]x_i\partial_j=-\rho e^{-i\varphi} z_{ij}^{(+)}\\
  \intertext{and}
 a_{21}=a_{43}&= [2x_1x_2 + i(x_2^2-x_1^2) ]\partial_2+ [(x_1^2-x_2^2)+2ix_1x_2  ]\partial_1
 \nonumber
 \\
 &= [x_1x_2 + ix_2^2+x_1x_2-ix_1^2 ]\partial_2+ [ix_1x_2+x_1^2+ix_1 x_2-x_2^2  ]\partial_1
 \nonumber
 \\
 &= [x_2(x_1+ix_2)-ix_1(x_1+ix_2) ]\partial_2+ [x_1(x_1+ix_2)+x_2(x_1+ix_2)  ]\partial_1
 \nonumber
 \\
 &=\rho e^{i\varphi} [(x_2 - ix_1)\partial_2+(x_1+ix_2)\partial_1 ]
  \nonumber
 \\
 &=-\rho e^{i\varphi} [i\epsilon_{ij}x_i\partial_j-\delta_{ij}x_i\partial_j ]=-\rho e^{i\varphi} z_{ij}^{(-)},
 \end{align}
  where $z_{ij}^{(\pm)}=(i\epsilon_{ij}\pm \delta_{ij})x_i\partial_j$.

\section{Action of  the $z_{ij}^{(\pm)}$ operator on a spinor\label{appC}}
  It is interesting now to see how the $z_{ij}^{(\pm)}$ operator acts  on a function
  that depends on $\varphi$ and $\rho$. In the context of this work, the following are some appropriate cases:
  \begin{align}  
   \partial_j(e^{i\ell\varphi}J_{|\ell-\delta|})&=\frac{-i\ell\epsilon_{jm}x_m}{\rho^2}e^{i\ell\varphi}J_{|\ell-\delta|} +
   e^{i\ell\varphi}J_{|\ell-\delta|}^\prime p^\perp\frac{x_j}{\rho},\\[4pt]
  \partial_j \rho&=\frac{x_j}{\rho}, \\[4pt]
  \partial_j \varphi&=\frac{-\epsilon_{jk}x_k}{\rho^2},\\[4pt] 
  \partial_j(e^{i(\ell+1)\varphi}J_{|\ell-\delta|+\epsilon_\ell})&=
   \left[\frac{-i(\ell+1)\epsilon_{jm}x_m}{\rho^2}J_{|\ell-\delta|+\epsilon_\ell} +
   J_{|\ell-\delta|+\epsilon_\ell}^\prime p^\perp\frac{x_j}{\rho}\right]e^{i(\ell+1)\varphi}.
  \end{align}
 The product of the two last matrices in Eq.~(\ref{Hint4}) yields the following nonzero matrix elements:
     \begin{align} 
   a_{11}={}&i\epsilon_\ell \surd{E_{p-}} \surd{s_{-}}\frac{e^{-i\varphi}}{\rho^2}(i\epsilon_{ij}+\delta_{ij})x_i\partial_j [e^{i(\ell+1)\varphi}J_{\nu_2}(p^\perp \rho) ]
   \nonumber
   \\[3pt]
   ={}&i\epsilon_\ell \surd{E_{p-}} \surd{s_{-}}\frac{e^{-i\varphi}}{\rho^2}(i\epsilon_{ij}+\delta_{ij}) \left[\frac{-i(\ell+1)}{\rho^2}\epsilon_{jm}x_ix_m J_{\nu_2}+J_{\nu_2}^\prime
  p^\perp \frac{x_ix_j}{\rho} \right ]e^{i(\ell+1)\varphi}
  \nonumber
  \\[3pt]
  ={}&i\epsilon_\ell \surd{E_{p-}} \surd{s_{-}}\frac{1}{\rho^2} [(\ell+1)J_{\nu_2}+J_{\nu_2}^\prime p^\perp \rho  ]e^{i\ell\varphi},\\[6pt] 
   a_{21}={}&\surd{E_{p+}} \surd{s_{+}}\frac{e^{i\varphi}}{\rho^2}(i\epsilon_{ij}-\delta_{ij})x_i\partial_j [e^{i\ell\varphi}J_{\nu_1}(p^\perp \rho) ]
   \nonumber
   \\[3pt]
   ={}& \surd{E_{p+}} \surd{s_{+}}\frac{e^{i\varphi}}{\rho^2}(i\epsilon_{ij}-\delta_{ij})\left[\frac{-i\ell\epsilon_{jm}}{\rho^2}x_ix_m J_{\nu_1}+J_{\nu_1}^\prime p^\perp\frac{x_ix_j}{\rho} \right]e^{i\ell\varphi}
   \nonumber
  \\[3pt]
  ={}&\surd{E_{p+}} \surd{s_{+}}\frac{1}{\rho^2} [\ell J_{\nu_1}-J_{\nu_1}^\prime p^\perp \rho ]e^{i(\ell+1)\varphi},
 \\[6pt]
   a_{31}={}&i\surd{E_{p-}} \surd{s_{+}}\frac{e^{-i\varphi}}{\rho^2}(i\epsilon_{ij}+\delta_{ij}) \left[\frac{-i(\ell+1)}{\rho^2}\epsilon_{jm}x_ix_m J_{\nu_2}+J_{\nu_2}^\prime p^\perp \frac{x_ix_j}{\rho} \right]e^{i(\ell+1)\varphi}
   \nonumber
   \\[3pt]
   ={}&i\frac{\surd{E_{p-}} \surd{s_{+}}}{\rho^2} [(\ell+1)J_{\nu_2}+J_{\nu_2}^\prime p^\perp \rho ]e^{i(\ell+1)\varphi}, 
 \\[6pt]
   a_{41}={}&\surd{E_{p+}} \surd{s_{-}}\frac{e^{i\varphi}}{\rho^2}(i\epsilon_{ij}-\delta_{ij}) \left[\frac{-i\ell}{\rho^2}\epsilon_{jm}x_mx_i J_{\nu_1}+J_{\nu_1}^\prime p^\perp \frac{x_ix_j}{\rho} \right]e^{i\ell\varphi}
  \\[3pt]
   ={}&\frac{\surd{E_{p+}} \surd{s_{-}}}{\rho^2} [\ell J_{\nu_1}-J_{\nu_1}^\prime p^\perp \rho ]e^{i(\ell+1)\varphi},
  \end{align}
 where $\epsilon_{ij}\epsilon_{jm}x_ix_j=\rho^2$ and $\epsilon_{jm}\delta_{ij}x_ix_m=\epsilon_{ij}x_ix_j=0$.

\end{document}